# A NEW APPROACH TO STATISTICAL ANALYSIS OF ELECTION RESULTS

**Ivan H. Krykun***

Vasyl' Stus Donetsk National University

& Institute of Applied Mathematics and Mechanics of NAS of Ukraine

**Abstract**

In this paper, a new method of detection of election fraud is proposed. This method is based on the calculation of the ratio of two standard normal random variables; estimation of parameters of obtained sample and comparison of these estimates with known theoretical values of parameters. Also in the paper, there is an example of the application of the method.

## 1. Introduction.

Elections are an important and interesting example of the practical implementation of the random behavior of human society. After all, the results of elections depend not only on, generally speaking, random and variable support for candidates, but also to a certain extent on the – also random – turnout of voters, which, in the case of approximately equal support for candidates, becomes decisive.

---

* *e-mail:* iwanko@i.ua

In fact, election processes in many countries, Ukraine and the Russian Federation in particular show that electoral fraud can also have a noticeable or (in the case of the Russian Federation) decisive influence on the election results – from relatively legal (such technologies as "black PR", motivation "own" voters to visit the polling stations, demotivating "other" voters or presenting "own" candidate as the "lesser evil" for those voters who are undecided) to direct falsifications (casting ballots for "own" candidates, spoiling ballots for candidates rivals, falsification of final protocols).

The implementation of legal election technologies (the motivation of "own" voters or demotivation of "other" voters through social networks and contextual political advertising as a tool for such motivation in particular) could be observed in recent election processes in such stable democracies as Great Britain (the 2016 United Kingdom European Union withdrawal membership referendum; the technologies used on this referendum are described in the film "Brexit: The Uncivil War") and the USA (the 2016 United States presidential election when Donald Trump was elected President; the technologies used on these both campaign are described in the film "The Great Hack").

Direct falsifications of election results could be observed, according to the conclusions of international independent observers, in several elections in Ukraine (in particular, the presidential elections of 2004, when massive falsifications (OSCE 2004) in favor of the pro-government candidate Victor Yanukovych led to mass national protests, which were called "Orange Revolution" and the re-voting of the decisive II round, which led to the victory of the opposition candidate Victor Yushchenko) and in fact in all election processes in the Russian Federation for at least the last 20 years.

Several approaches have been proposed by researchers to detect possible electoral fraud (Myagkov 2009).

First, the authors look at the frequency distribution of turnout or candidate's results across electoral districts. If the data are genuine, we would typically expect this distribution to be normal. Deviations from normality – such as bimodality or a substantial tail of districts with very high turnouts – give cause for suspicion. Also on such frequency distribution of turnout is noticeable increasing of round numbers (such as, e.g., multiples of 5 or 10) (Kobak et al. 2016a).

Second, the authors look at how each candidate's share of the total electorate correlates with the turnout. If voting figures are reported accurately, we would expect a candidate's share of the electorate to increase with turnout in rough proportion to his or her overall vote share: if turnout increases by 100 votes, we would expect a candidate who scores 50% of the vote overall to pick up around 50 of these votes. If, by contrast, turnout is inflated by the casting of ballot boxes, the proportion of additional votes captured by the favored candidate will be far higher.

The third approach is the analysis of the ratio of votes cast by candidates/parties, depending on the turnout. It was first used by J. B. Kiesling (2004) to analyze elections in Armenia. The method has proven itself useful in Russia to identify fraud in favor of the ruling party and its candidates (Kobak et al. 2016b: Rozenas 2017).

Another approach (Deckert et al. 2011; Nigrini 2012) is the use of Benford's law (or the first-digit law) to compare the obtained frequency of appearance of the first (or second) digit in the final election documentation with the theoretical distribution.

A more detailed review of statistical approaches can be found in Myagkov (2009).

However, all proposed statistical approaches to the assessment of election fraud are subject to legitimate criticism, because the deviation of the empirical frequency distribution from the normal distribution can be caused by other reasons (for example, regional differences in voter activity or support for certain candidates).

This paper proposes a new approach to assessing the presence of election fraud. The main idea is to construct, after certain processing of statistical data, an arctangent regression of the obtained data and estimate the parameters of this arctangent regression as parameters of the Cauchy distribution according to the results Krykun (2020a,b).

## 2. Main result

We recall (Walck 1996; Krishnamoorthy 2006) that the Cauchy distribution with parameters $\alpha$ and $\gamma$ (where $\gamma > 0$) has probability density function

$$f(x) = \frac{1}{\pi} \frac{\gamma}{(x-\alpha)^2 + \gamma^2}, \quad \text{for } -\infty < x < +\infty,$$

and cumulative distribution function, respectively,

$$F(x) = \frac{1}{2} + \frac{1}{\pi} \arctan \frac{x-\alpha}{\gamma}.$$

The parameters $\alpha$ and $\gamma$ are called, respectively, a location parameter and a scale one.

It is well known (Walck 1996; Krishnamoorthy 2006) that the Cauchy distribution has neither mathematical expectations nor moments of higher orders. The Cauchy distribution belongs to the distributions with "heavy tails", for which the law of large numbers does not hold. Due to these properties of the Cauchy distribution, it is not possible to estimate its parameters by standard methods (method of moments or maximum likelihood estimation), several different methods have been proposed for this.

Based on the fact that the cumulative distribution function for the Cauchy distribution is the arctangent function, Krykun (2020b, Prop. 2) proposed a simple approach to estimate the unknown parameters of the Cauchy distribution, based on the use of empirical arctangent regression.

One more well-known result (Walck 1996; Krishnamoorthy 2006) is that the ratio between two standard normal random variables with parameters 0 and 1 is a random variable having a Cauchy distribution with parameters 0 and 1.

On the other hand, many electoral indicators (turnout, percentage of votes for certain candidates, percentage of spoiled ballots or votes "against all") are plausible to consider as having random deviations distributed according to a normal distribution. Thus, if we find the mathematical expectation and variance of these quantities, we can normalize them.

**Proposition 1.** After normalizing two selected election indicators, we calculate their ratio. Next, we will perform statistical processing of the obtained data according to Krykun (2020b, Prop. 3 and Lemma 1) and obtain empirical estimates of the distribution parameters. Further, we will compare the obtained estimates with the theoretical parameters of the Cauchy distribution (that is, with the values 0 and 1) and, in case of noticeable differences, try to estimate the probability of such a deviation of the parameters from the theoretical ones and draw statistical conclusions about the plausibility of the election result.

**Remark 1.** In order to see the picture better, it is advisable to separate a set of polling stations, if there exists evidence of mass violations of the same type (for example, casting ballots for "own" candidate or spoiling ballots for an "other" candidates) and analyze a set of data from these stations.

**Remark 2.** If there are reasons to believe that the increase in turnout was massively falsified (because of casting in ballots or multiple votes), then it is advisable to calculate the final election indicators (turnout to poll, votes against all and invalid ballots) without data from such electorates precincts.

## 3. An example. Analysis of the results of the 1st round of 2004 Ukrainian presidential elections in Donetsk and Luhansk regions.

3.1. Background.

During the Ukrainian presidential elections in 2004, in Donetsk region were formed 23 constituencies and 13 constituencies were formed in the Luhansk region – a total of 35 constituencies (from the total number of 225 constituencies and foreign electoral district). Donetsk and Luhansk regions in those elections were the basis for the pro-government candidate Victor Yanukovych, who in these regions won 39,5% of all votes in his support, namely 4,35 million out of 11,0 million, according to the results of the first round (CEC Ukraine 2004).

At these elections, observers recorded mass violations in favor of the government-supported candidate, the notorious political figure Victor Yanukovych. These violations were also recorded during the 1st round, and especially during the 2nd round of elections, which led to mass protests in November-December 2004 (Orange Revolution) and the re-voting of the decisive 2nd round, which took place 12/26/04. The falsifications were especially large-scale in Donetsk and Luhansk regions, which is clearly evidenced by the statistical results of these three rounds of voting (here PEC = precinct election commission):

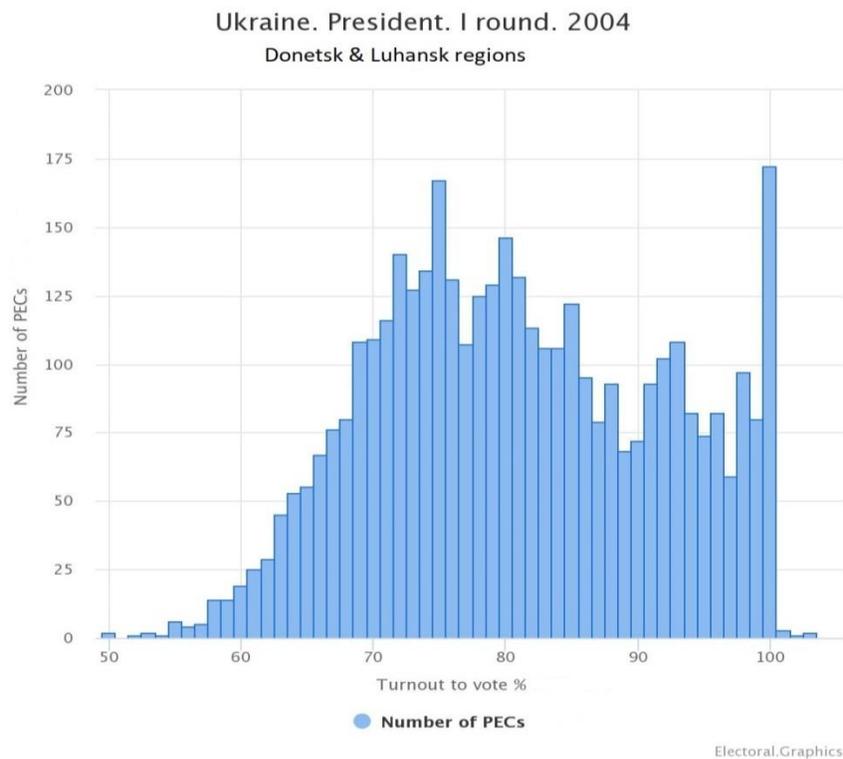

Figure 1. Turnout to vote and number of PECs on 1st round (Graphical data 2004)

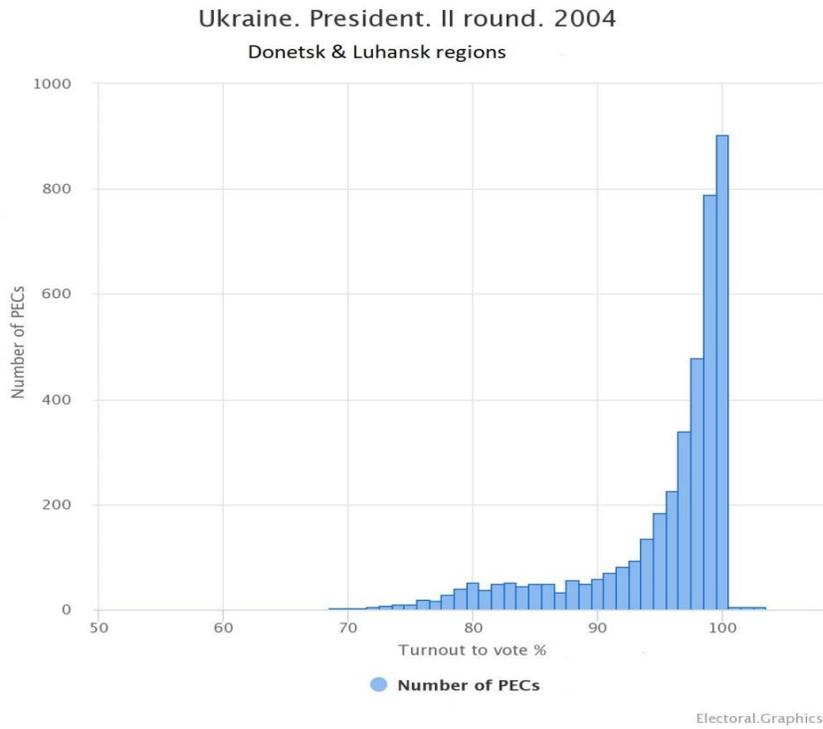

Figure 2. Turnout to vote and number of PECs on 2nd round (Graphical data 2004)

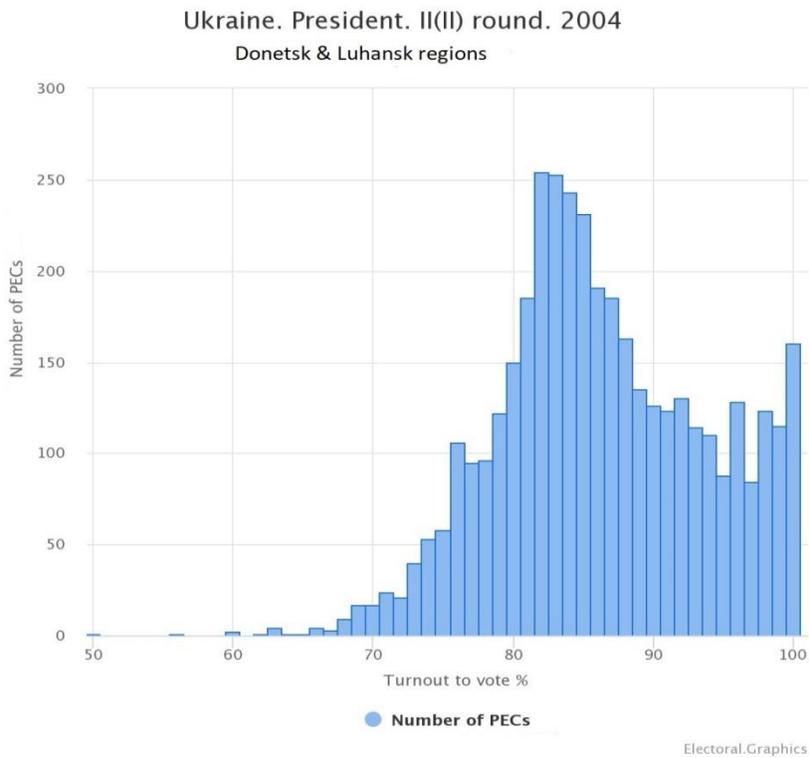

Figure 3. Turnout to vote and number of PECs on revoting of 2nd round (Graphical data 2004)

Moreover the following table is indicative (CEC Ukraine 2004):

Table 1. Turnout in three rounds of that election.

|  | 1st round, number of registered voters | Turnout, voters and % from the 1st round list | | |
|---|---|---|---|---|
|  |  | 1st round | 2nd round | Re-voting of 2nd round |
| Donetsk region | 3,685 million | 2,878 million 78,1 % | 3,712 million **100,7 %** | 3,144 million 85,3 % |
| Luhansk region | 1,946 million | 1,472 million 75,6 % | 1,754 million 90,1 % | 1,638 million 84,2 % |

According to these histograms, we have absolutely impossible turnout on the 2nd round (indicates total falsifications at the PEC), we have a bimodal distribution on the histogram of the 1st round and several turnout indicators in excess of 100% (it indicates massive falsifications at the PEC) and a fairly plausible picture on re-voting of the 2nd round (a certain overestimation of the number of PECs with a high turnout is caused by the presence of special precincts – hospitals, military units, prisons with a small number of voters and a high turnout rate).

The fantastic turnout during the 2nd round of elections in the entire Donetsk region of more than 100% (from the number of voters on the list of the 1st round) only confirms not just the mass, but the totality of falsifications!

3.2. Estimation of parameters

Therefore, in the 1st round, we can assume the presence of falsifications associated with an increase in voter turnout (due to multiple voting or direct ballot casting). Therefore, we will choose 2 indicators for the analysis – the percentage of voter turnout (which may be overstated) and the percentage of votes against all (which may be understated, since the falsifiers probably added ballots for "their" candidate, and not against all).

Further, in order to avoid the influence of inflated turnout, we will calculate the statistical indicators of the 1st round of this election without the votes from Donetsk and Luhansk regions. We have (calculations by the author according to official data (CEC Ukraine 2004; Dataset 2004)):

Table 2. Mean and variance of turnout and number of votes against all on 1$^{st}$ round of that election without the votes from Donetsk and Luhansk regions.

|          | Turnout       | Votes against all |
|----------|---------------|-------------------|
| Mean     | 74,502 %      | 2,027 %           |
| Variance | 46,376 sq. %  | 1,857 sq. %       |
| $\sigma$ | 6,810 %       | 1,363 %           |

Next, we normalize and center the final election indicators of turnout and votes against all 35 constituencies in Donetsk and Luhansk regions:

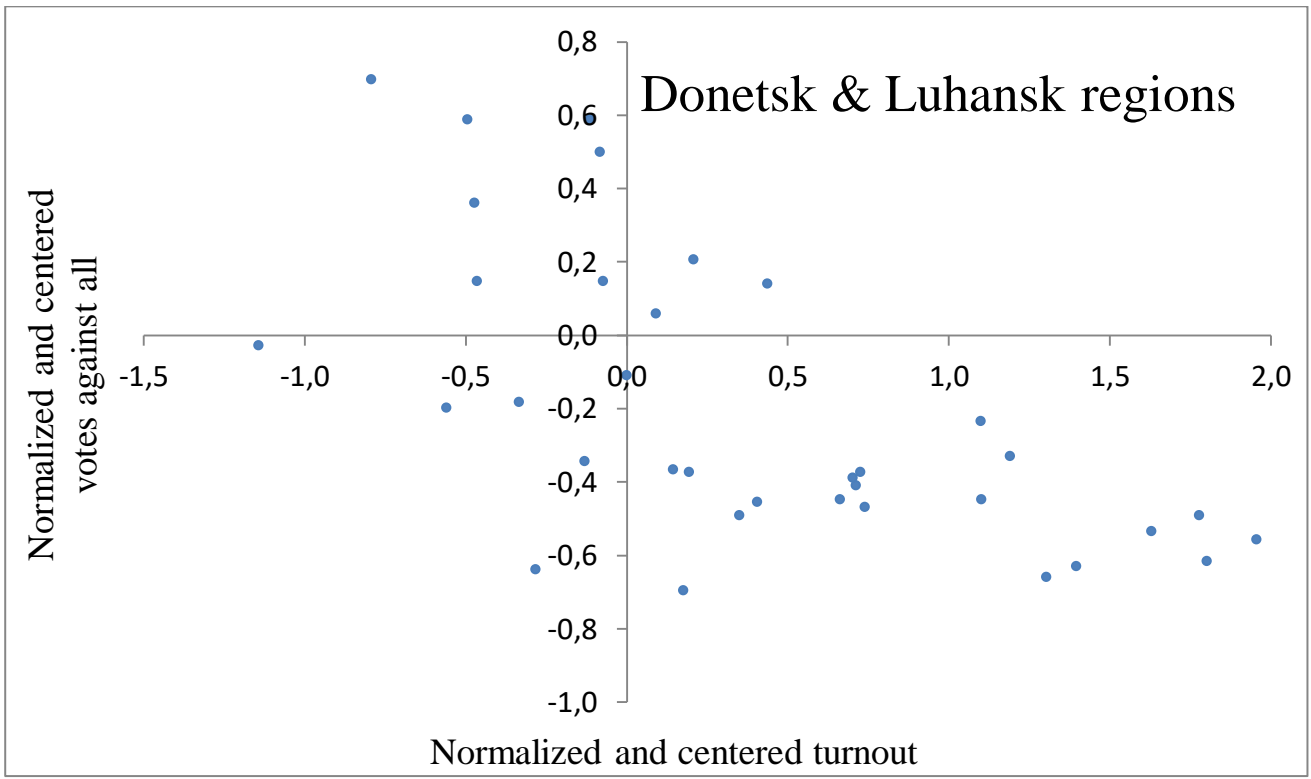

Figure 4. Normalized and centered turnout and votes against all.

Later we calculate their ratio according to the formula

$$x_i = \frac{\text{normalized and centered \% of turnout in the } i\text{th constituency}}{\text{normalized and centered \% of votes against all in the } i\text{th constituency}}.$$

For the obtained sample $x_i$, we apply the procedure (Krykun 2020b, Prop. 2) and use (Krykun 2020b, eq. (10), (11)). In addition, for reasons (Krykun 2020b, Prop. 3) we reject some quantity of largest and smallest values from the sample $x_i$.

After the specified data processing, we will get a table of parameter estimates:

Table 3. Estimates of parameters.

| Estimate \ Reject | 1 items | 3 items | 7 items | 9 items |
|---|---|---|---|---|
| $\alpha$ | -0,9371 | -1,0211 | -1,0366 | -1,0281 |
| $\gamma$ | 0,7564 | 1,0249 | 1,1813 | 1,2596 |

So, we can see that the point estimate of the location parameter $\alpha$ is quite stable and is within $[-1,04; -1,02]$, and the point estimate of the scale parameter $\gamma$ increases with decreasing sample size and is within $[1,02; 1,26]$.

### 3.3. Plausibility of the election results

Therefore, the parameter is $\gamma$ close to the theoretical value equal to 1, but the parameter α is significantly different from the theoretical value equal to 0.

Let's consider a sample, which consists of observations of the realization of a random variable distributed by the Cauchy law with an unknown parameter $\alpha$ and a known parameter $\gamma$. We know the mean of the sample $\bar{x} = 0,1965$ and suppose for simplicity, that the parameter $\gamma$ is known. Further, we calculate the theoretical probability that the parameter $\alpha$ is contained in the extended, compare to the obtained estimates, interval, for example in the interval $[-1,1; -0,95]$.

For this purpose we will use such property of the Cauchy distribution (Walck 1996; Krishnamoorthy 2006): for the sample $\{x_i\}$, distributed by the Cauchy law with the parameters $\alpha$ and $\gamma$, the variable $\bar{x} = \frac{1}{n}\sum_{i=1}^{n} x_i$ is also distributed by the Cauchy law with the same parameters.

So we obtain (for $\gamma = 1$)

$$P\{\alpha \in [-1,1; -0,95]\} = P\{-1,1 \leq \alpha \leq -0,95\} =$$

$$= P\{-1,1 - \bar{x} \leq \alpha - \bar{x} \leq -0,95 - \bar{x}\} =$$

$$= \frac{1}{\pi}\arctan\frac{-0,95 - \bar{x}}{\gamma} - \frac{1}{\pi}\arctan\frac{-1,1 - \bar{x}}{\gamma} =$$

$$= \frac{1}{\pi}\arctan(-1,1465) - \frac{1}{\pi}\arctan(-1,2965) = 0,0192.$$

Let's compare result above with result for $\gamma = 1,26$:

$$P\{\alpha \in [-1,1; -0,95]\} =$$

$$= \frac{1}{\pi}\arctan\left(\frac{1,2965}{1,26}\right) - \frac{1}{\pi}\arctan\left(\frac{1,1465}{1,26}\right) = 0,0195.$$

And two more probabilities (for $\gamma = 1$):

$$P\{\alpha \in [-1,04; -1,02]\} = 0,0025,$$

and for $\gamma = 1,26$

$$P\{\alpha \in [-1,04; -1,02]\} = 0,0026.$$

Therefore, one can say that the results in Donetsk and Luhansk regions during the 1st round of these elections are extremely unlikely, which gives ground for asserting the presence of significant fraud.

## 4. General conclusions and prospects for further research

The main advantages of the proposed method are the possibility of using such indicators of an election that are difficult for falsifiers to predict, such as the number of spoiled ballots, the number of votes "against all" and taking into account such a statistical indicator as the variance of election data. It is also possible to use the usual statistical indicators – turnout, and support for certain candidates.

The author assumes that with a rational (that is, taking into account the specifics of possible violations) pairwise analysis of the mentioned indicators of the election process, a noticeable deviation of the distribution of the ratio of these values from the Cauchy distribution with parameters 0 and 1 gives reason to conclude the presence of election fraud.

In addition, for a numerical assessment of the influence of falsifications on the election results, the author supposes it appropriate to use the Cauchy distribution curve instead of a normal distribution one, because the Cauchy distribution as a distribution with "heavy tails", can better take into account the specifics of election processes (for example, close to 100 % turnout in special precincts – hospitals, prisons, military units, or greatly increased support for the candidate in his hometown).

## **Acknowledgments.**

The author expresses his heartfelt gratitude to the brave soldiers of the Ukrainian army who protect the lives of the author and his family from Russian bloody murderers since 2014.

The author also gratefully acknowledges the project "Electoral Memory" (ukr.vote) of the public organization "Ukrainian Center for Social Data" (https://socialdata.org.ua) and personally Mr. Serhij Vasylchenko, as well as Mr. Roman Udot, Co-chairman of the Board of The Movement for the Defense of Voters' Rights "Golos" for the help with the finding of statistical data.
## **References**

1. Dataset of 2004 Ukrainian presidential election results. (2004) Available at https://www.electoral.graphics/Portals/0/EasyDNNnews/Uploads/125/2004.10.31%20President%20Ukraine.zip

2. Deckert, J., Myagkov, M. and Ordeshook, P. C. (2011) "Benford's Law and the Detection of Election Fraud." *Political Analysis*, 19, No. 3, 245-268.

3. Graphical data. (2004) Available at https://www.electoral.graphics/ru-ru/histogram-generator

4. Kiesling, J. B. (2004) *Charting Electoral Fraud: Turnout Distribution Analysis as a Tool for Election Assessment*.